\documentstyle[epsfig,12pt]{article}
\def\1ad{\mbox{\normalsize $^1$}}
\def\2ad{\mbox{\normalsize $^2$}}
\def\3ad{\mbox{\normalsize $^3$}}
\def\4ad{\mbox{\normalsize $^4$}}
\def\5ad{\mbox{\normalsize $^5$}}
\def\6ad{\mbox{\normalsize $^6$}}
\def\7ad{\mbox{\normalsize $^7$}}
\def\8ad{\mbox{\normalsize $^8$}}
\def\makefront{\vspace*{1cm}\begin{center}
\def\newtitleline{\\ \vskip 5pt}
{\Large\bf\titleline}\\
\vskip 1truecm
{\large\bf\authors}\\
\vskip 5truemm
\addresses
\end{center}
\vskip 1truecm
{\bf Abstract:}
\abstracttext
\vskip 1truecm}
\newcommand{\be}{\begin{equation}}                                              
\newcommand{\ee}{\end{equation}}                                                
\newcommand{\half}{\frac{1}{2}}

\newcommand{\LCB}{\raisebox{-0.3ex}{\mbox{\LARGE$\left\{\right.$}}}
\newcommand{\RCB}{\raisebox{-0.3ex}{\mbox{\LARGE$\left.\right\}$}}}

\begin {document}
\def\titleline{
 SYM on the lattice
}
\def\authors{
 Istv\'an Montvay
}
\def\addresses{
 Deutsches Elektronen Synchrotron DESY, \\
 Notkestr.~85, D-22603 Hamburg, Germany
}
\def\abstracttext{
 Non-perturbative predictions and numerical simulations in
 supersymmetric Yang-Mills (SYM) theories are reviewed.
}
\makefront
\section{Introduction}                                      \label{sec1}

 The investigation of non-perturbative properties of supersymmetric
 (SUSY) gauge theories is presently a rather theoretical subject: it is
 not known whether (broken) supersymmetry is realized in nature or
 not.
 Moreover, the simple (``minimal'') supersymmetric extensions of the
 standard model do not involve strong interactions near or above the
 scale of supersymmetry breaking, hence the knowledge of the
 non-perturbative supersymmetric dynamics is not directly required.
 In spite of this there is a continuing interest in studying strongly
 interacting quantum field theories with supersymmetry.
 (For an early review of the subject see ref.~\cite{AKMRV}.)

 The motivation to investigate non-perturbative features of
 supersymmetric gauge theories is partly coming from the desire to
 understand relativistic quantum field theories better in general:
 the supersymmetric points in the parameter space of all quantum field
 theories are very special since they correspond to situations of a
 high degree of symmetry.
 As recent work of Seiberg and Witten \cite{SEIWIT} and other related
 papers showed, there is a possibility to approach non-perturbative
 questions in four dimensional quantum field theories by starting from
 exact solutions in some highly symmetric points and treat the symmetry
 breaking as a small perturbation.
 Beyond this, the knowledge of non-perturbative dynamics in
 supersymmetric quantum field theories can also be helpful in 
 understanding the greatest puzzle of the standard model, with or
 without supersymmetric extensions, namely the existence of a large
 number of seemingly free parameters.
 As we know from QCD, strong interactions in non-abelian gauge theories
 are capable to reproduce from a small number of input parameters a
 large number of dynamically generated parameters for quantities
 characterizing bound states.
 This is a possible solution also for the parameters of the standard
 model if new strong interactions are active beyond the electroweak
 symmetry breaking scale.

\subsection{$N=1$ SYM}                                    \label{sec1.1}

 The simplest supersymmetric gauge theory is the supersymmetric
 extension of Yang-Mills theory.
 The action of Yang-Mills theory with $N=1$ supersymmetry is
 conventionally given as
$$
\int d^4x\, d^2\theta\, {\rm Tr}(W^\alpha W_\alpha)
$$
\be\label{eq01}
= \int d^4x\, {\rm Tr}\left\{ -\half F_{\mu\nu}F^{\mu\nu}
+{i \over 2} F_{\mu\nu}\tilde{F}^{\mu\nu}
-i\lambda\sigma^\mu(D_\mu\bar{\lambda})
+i(D_\mu\bar{\lambda})\bar{\sigma}^\mu\lambda + D^2
\right\} \ ,
\ee
 where the first line is written in terms of the spinorial field
 strength superfield $W(x,\theta,\bar{\theta})_\alpha$ which depends on
 the four-coordinate $x$ and the anticommuting Weyl-spinor variables
 $\theta_\alpha,\bar{\theta}_{\dot\alpha}$ ($\alpha,\dot{\alpha}=1,2$).
 After performing the Grassmannian integration on $\theta$, one obtains
 the second form in terms of the component fields.
 The field strength tensor $F_{\mu\nu}$ and its dual are defined, as
 usual, as
\be\label{eq02}
F_{\mu\nu}(x) \equiv -ig F_{\mu\nu}^r(x) T_r \ , \hspace{3em}
\tilde{F}_{\mu\nu} \equiv \half\epsilon_{\mu\nu\rho\sigma} 
F^{\rho\sigma} \ .
\ee
 $\lambda,\bar{\lambda}$ represent a Majorana fermion field in the
 adjoint representation and $D$ is an auxiliary field.

 The action in (\ref{eq01}) includes a $\Theta$-term, therefore it
 is natural to introduce the complex coupling
\be\label{eq03}
\tau \equiv \frac{\Theta}{2\pi} + \frac{4\pi i}{g^2}
\ee
 and then, with arbitrary $\Theta$, the $N=1$ SYM action becomes:
$$
\frac{1}{4\pi} \Im \left\{ \tau
\int d^4x\, d^2\theta\, {\rm Tr}(W^\alpha W_\alpha) \right\} 
$$
$$
= \frac{1}{g^2}\int d^4x\, {\rm Tr}\left[ -\half F_{\mu\nu}F^{\mu\nu}
-i\lambda\sigma^\mu(D_\mu\bar{\lambda})
+i(D_\mu\bar{\lambda})\bar{\sigma}^\mu\lambda + D^2
\right]
$$
\be\label{eq04}
+\frac{\Theta}{16\pi^2} \int d^4x\, 
{\rm Tr} \left[ F_{\mu\nu}\tilde{F}^{\mu\nu} \right] \ .
\ee

 The remarkable feature of the action in (\ref{eq04}) is that, after
 performing the trivial Gaussian integration over the auxiliary field
 $D$, it is nothing else than an ordinary Yang-Mills action with a
 massless Majorana fermion in the adjoint representation.
 This shows that this theory is ``automatically'' supersymmetric.
 Introducing a non-zero gaugino mass $m_{\tilde{g}}$ breaks
 supersymmetry ``softly''.
 Such a mass term is:
\be\label{eq05}
m_{\tilde{g}} (\lambda^\alpha\lambda_\alpha + 
\bar{\lambda}^{\dot{\alpha}} \bar{\lambda}_{\dot{\alpha}} )
= m_{\tilde{g}} (\bar{\Psi}\Psi) \ .
\ee
 Here in the first form the Majorana-Weyl components
 $\lambda,\bar{\lambda}$ are used, in the second form the Dirac-Majorana
 field $\Psi$.

 The Yang-Mills theory of a Majorana fermion in the adjoint
 representation is, in a general sense, similar to QCD: besides the
 special Majorana-feature the only difference is that the fermion is in
 the adjoint representation and not in the fundamental one.
 As in QCD, a central feature of low-energy dynamics is the realization
 of the global chiral symmetry.
 As there is only a single Majorana adjoint ``flavour'', the global
 chiral symmetry of $N=1$ SYM is $U(1)_\lambda$, which coincides with
 the so called {\em R-symmetry} generated by the transformations
\be\label{eq06}
\theta_\alpha^\prime = e^{i\varphi}\theta_\alpha \ , \hspace{2em}
\bar{\theta}_{\dot{\alpha}}^\prime = 
e^{-i\varphi}\bar{\theta}_{\dot{\alpha}} \ .
\ee
 This is equivalent to
\be\label{eq07}
\lambda_\alpha^\prime = e^{i\varphi}\lambda_\alpha \ , \hspace{0.5em}
\bar{\lambda}_{\dot{\alpha}}^\prime = 
e^{-i\varphi}\bar{\lambda}_{\dot{\alpha}} \ , \hspace{0.5em}
\Psi^\prime = e^{-i\varphi\gamma_5}\Psi \ .
\ee

 The $U(1)_\lambda$-symmetry is anomalous: for the corresponding axial
 current $J_\mu \equiv \bar{\Psi}\gamma_\mu\gamma_5\Psi$, in case of
 $SU(N_c)$  gauge group with coupling $g$, we have
\be\label{eq08}
\partial^\mu J_\mu = \frac{N_c g^2}{32\pi^2} 
\epsilon^{\mu\nu\rho\sigma} F_{\mu\nu}^r F_{\rho\sigma}^r \ .
\ee
 However, the anomaly leaves a $Z_{2N_c}$ subgroup of $U(1)_\lambda$
 unbroken.
 This can be seen, for instance, by noting that the transformations
\be\label{eq09}
\Psi \to e^{-i\varphi\gamma_5}\Psi \ , \hspace{2em}
\bar{\Psi} \to \bar{\Psi}e^{-i\varphi\gamma_5}
\ee
 are equivalent to
\be\label{eq10}
m_{\tilde{g}} \to m_{\tilde{g}} e^{-2i\varphi\gamma_5} \ ,\;
\Theta_{SYM} \to \Theta_{SYM} - 2N_c\varphi \ ,
\ee
 where $\Theta_{SYM}$ is the $\Theta$-parameter of gauge dynamics.
 Since $\Theta_{SYM}$ is periodic with period $2\pi$, for
 $m_{\tilde{g}}=0$ the $U(1)_\lambda$ symmetry is unbroken if
\be\label{eq11}
\varphi = \varphi_k \equiv \frac{k\pi}{N_c} \ , \hspace{2em}
(k=0,1,\ldots,2N_c-1) \ .
\ee
 For this statement it is essential that the topological charge is
 integer.

 The discrete global chiral symmetry $Z_{2N_c}$ is expected to be 
 spontaneously broken by the non-zero {\em gaugino condensate}
 $\langle \lambda\lambda \rangle \ne 0$ to $Z_2$ defined by 
 $\{\varphi_0,\varphi_{N_c}\}$ (note that $\lambda \to -\lambda$ is a
 rotation).
 The consequence of this spontaneous chiral symmetry breaking pattern
 is the existence of a first order phase transition at zero gaugino
 mass $m_{\tilde{g}}=0$.
 For instance, in case of $N_c=2$ there exist two degenerate ground
 states with opposite signs of the gaugino condensate.
 The symmetry breaking is linear in  $m_{\tilde{g}}$, therefore the
 two ground states are exchanged at $m_{\tilde{g}}=0$ and there is a
 first order phase transition.

\begin{figure}
\begin{center}
\epsfig{file=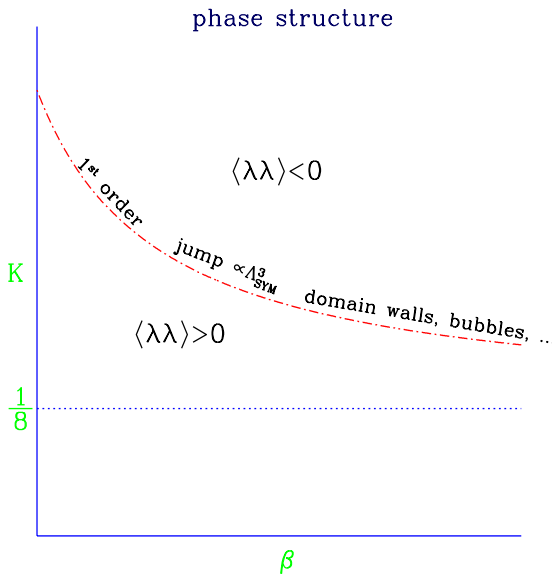,
        width=8.0cm,height=8.0cm,
        bbllx=80pt,bblly=190pt,bburx=680pt,bbury=760pt,
        angle=0}
\end{center}
\vspace{-4.0em}
\begin{center}
\parbox{13cm}{\caption{ \label{fig1}
 Expected phase structure of SU(2) Yang-Mills theory with a Majorana
 fermion in adjoint representation in the ($\beta,K$)-plane.
 The dashed-dotted line is a first order phase transition at zero gluino
 mass, where supersymmetry is expected.}}
\end{center}
\end{figure}

 The non-perturbative features of the SYM theory can be investigated in
 a lattice formulation.
 As always, the lattice action is not unique (see section \ref{sec3.1}).
 A possible formulation was given by Curci and Veneziano \cite{CURVEN}
 based on the well known lattice formulation of QCD introduced by
 Wilson.
 In the lattice action the {\em bare gauge coupling} (of $SU(N_c)$)
 is convetionally represented by $\beta \equiv 2N_c/g^2$ and the
 {\em bare gaugino mass} by the {\em hopping parameter} $K$.
 In the plane of ($\beta,K$) there is a {\em critical line}
 corresponding to zero gaugino mass and the expected phase structure is
 the one shown in figure~\ref{fig1}.

\subsection{$N=2$ SYM}                                    \label{sec1.2}

 The SYM theory with $N=2$ {\em extended supersymmetry} is a highly
 constrained theory which has, however, more structure than the
 relatively simple $N=1$ case discussed above.
 In particular, besides the $N=1$ ``vector superfield'' containing the
 gauge boson and gaugino $(A_\mu,\lambda)$, it also involves an $N=1$
 ``chiral superfield'' $(\phi,\lambda^\prime)$ in the adjoint
 representation which consists of the complex scalar $\phi \equiv A+iB$
 and the Majorana fermion $\lambda^\prime$.
 The Majorana pair $(\lambda,\lambda^\prime)$ can be combined to a
 Dirac-fermion $\psi$ and then the vector-like (non-chiral) nature of
 this theory can be made explicit.

 The Euclidean action of $N=2$ SYM theory in component notation, for
 simplicity in case of an SU(2) gauge group, is the following:
$$
S_{SYM}^{N=2} = \int d^4x\;  \LCB 
{1 \over 4} F_{\mu\nu}^r(x)F_{\mu\nu}^r(x)
+ \half (D_\mu A^r(x))(D_\mu A^r(x)) 
+ \half (D_\mu B^r(x))(D_\mu B^r(x))
$$
$$
+ \overline{\psi}^r(x)\gamma_\mu D_\mu \psi^r(x) 
+ ig\epsilon_{rst} \overline{\psi}^r(x) 
[A^s(x) + i\gamma_5 B^s(x)] \psi^t(x)
$$
\be\label{eq12}
+ \frac{g^2}{2} [A^r(x)A^r(x)B^s(x)B^s(x) 
- A^r(x)B^r(x)A^s(x)B^s(x)]   \RCB  \ .
\ee
 This is a massless adjoint Higgs-Yukawa model with special Yukawa- and
 quartic couplings given in terms of the gauge coupling.

 In $N=2$ SUSY possible couplings are so strongly constrained by the
 symmetry that only gauge couplings are allowed.
 Another important feature is that the symmetry also implies that
 the matter field content is always vector-like.
 Therefore $N=2$ SUSY theories are always non-chiral and hence well
 suited for lattice sudies.
 A lattice formulation of $N=2$ SYM based on the Wilsonian formulation
 of QCD has been investigated in \cite{N=2}.

 The main new feature of $N=2$ SYM compared to $N=1$ SYM is that it also
 contains scalar fields, hence there is the possibility of Higgs
 mechanism.
 Let us here only consider the simplest case of an $SU(2)$ gauge group.
 In the Higgs phase the vacuum expectation value of the scalar field
 is non-zero.
 In terms of the real components $\phi(x) \equiv A(x)+iB(x)$ we have
\be\label{eq13}
\langle A^r(x) \rangle \ne 0 \ , \hspace{2em}
\langle B^r(x) \rangle \ne 0 \ .
\ee
 This implies the Higgs mechanism breaking of $SU(2) \to U(1)$,
 similarly to the Georgi-Glashow model.
 Due to the Higgs mechanism the ``charged'' gauge bosons become heavy.
 The low-energy effective theory is $N=2$ SYM with $U(1)$ gauge group.

 Seiber and Witten proved \cite{SEIWIT} that extended SUSY and
 asymptotic freedom can be exploited to determine exactly the low-energy
 effective action in the Higgs phase, if the vacuum expectation values
 are large.
 At strong couplings there are two singularities of the effective
 action corresponding to light monopoles and dyons, respectively.

 The expectation value of the complex scalar field
 $\langle \phi \rangle$ parametrizes the {\em moduli space} of
 zero-energy degenerate vacua.
 The degeneracy is a consequence of $N=2$ supersymmetry.
 This phenomenon is usually referred to as the existence of {\em flat
 directions}: the potential identically vanishes for $B^r=cA^r$.
 In the present case the moduli space is a non-compact manifold with
 two parameters.
 The presence of non-compact flat directions requires the breaking
 of supersymmetry for the definition of the path integral over the
 scalar fields: otherwise the path integral would be divergent.
 It is also plausible that soft breaking with mass terms is not enough
 in the Higgs phase, where the mass-squared terms in the potential are
 negative. 
 Therefore small hard breaking by dimensionless couplings is also
 required.

 These arguments are quite general.
 In the special case of $N=2$ SYM the general renormalizable scalar
 potential with the given set of scalar fields is
$$
V(A,B) \equiv \half m_A^2 A^rA^r +  \half m_B^2 B^rB^r 
$$
\be\label{eq14}
+ \lambda_A (A^rA^r)^2 + \lambda_B (B^rB^r)^2 
+ \lambda_{[AB]}A^rA^rB^sB^s - \lambda_{(AB)} (A^rB^r)^2 \ .
\ee
 N=2 supersymmetry is at the point of parameter space where
\be\label{eq15}
m_A=m_B=\lambda_A=\lambda_B=0 \ ,  \hspace{1em}
2\lambda_{[AB]}=2\lambda_{(AB)}=g^2 \ .
\ee
 In order that the path integral over the scalar fields be convergent, 
 the quartic couplings have to fulfil the following conditions:
$$
\lambda_A > 0  \hspace{2em} {\rm AND}  \hspace{2em}
\lambda_B > 0  \hspace{2em} {\rm AND} 
$$
\be\label{eq16}
\LCB \lambda_{[AB]} \geq \max(0,\lambda_{(AB)})
\hspace{2em}  {\rm OR}  \hspace{2em}
4\lambda_A\lambda_B > \max[\lambda_{[AB]}^2,
(\lambda_{[AB]} - \lambda_{(AB)})^2]  \RCB \ .
\ee
 This is in conflict with the supersymmetry conditions.

 The consequence of the conflict between supersymmetry and the
 convergence of the path integral over the scalar fields is that in
 a path integral formulation of the quantized theory the supersymmetry
 has to be broken.
 On the lattice this means that supersymmetry is broken as long as the
 lattice spacing is non-zero and can only be restored in the continuum
 limit $a \to 0$.

 The tuning to the supersymmetric point in $N=2$ SYM can be studied, for
 instance, in lattice perturbation theory \cite{N=2}.
 It can be shown that the compact flat direction is reproduced for
 $a \to 0$ on a specific phase transition where three different kinds of 
 Higgs-phases meet.
 The emergence of the non-compact flat direction is a result of
 cancelling of quantum correction contributions from scalars and
 fermions.
 This is similar to the situation which occurs if the so called vacuum
 stability boundary in Higgs-Yukawa models reaches zero fermion mass.
 (For a lattice investigation of the vacuum stability bound see, for
 instance, ref.~\cite{LMMPW}.)

\section{Non-perturbative predictions for $N=1$ SYM}        \label{sec2}

 In analogy with QCD, one expects that the spectrum of the SYM model
 consists of colourless bound states formed out of the fundamental
 excitations, namely gluons and gluinos.
 (In this context we shall use the name ``gluino'' instead of the more
 general term ``gaugino''.)
 In the supersymmetric point at zero gluino mass these bound states
 should be organized in supersymmetry multiplets, according to the
 representations of the SUSY extension of the Poincar\'e algebra.
 For the description of lowest energy bound states one can use an
 effective field theory in terms of suitably chosen colourless composite
 operators.
 
 For $N=1$ SYM the effective action was constructed by Veneziano and
 Yankielowicz (VY) \cite{VENYAN}.
 The composite operator appearing in the VY effective action is a
 chiral supermultiplet $S$ containing as component fields the
 expressions for the anomalies \cite{FERZUM}:
\be\label{eq17}
S \equiv A(y) + \sqrt{2} \theta\Psi(y) + \theta^2 F(y) \ ,
\ee
 where, for instance, the scalar component is proportional to the
 gluino bilinear
\be\label{eq18}
A \propto \lambda^\alpha \lambda_\alpha \ .
\ee
 The other components contain gluino-gluino and gluino-gluon
 combinations.
 Therefore, as far as a constituent picture is applicable to the bound
 states formed by strong interactions, the particle content of the
 lowest supersymmetry multiplet is: a pseudoscalar gluino-gluino bound
 state, a Majorana spinor gluon-gluino bound state and a scalar
 gluino-gluino bound state.
 In terms of $S$ the VY effective action has the form
\be\label{eq19}
S_{VY} = \frac{1}{\alpha}\int d^4x\, d^2\theta\, d^2\bar{\theta}\,
(S^\dagger S)^{1/3} +
\gamma\, [\int d^4x\, d^2\theta\, ( S \log \frac{S}{\Lambda} - S) 
+ {\rm h.c.}] \ .
\ee
 Here $\alpha$ and $\gamma$ are positive constants and $\Lambda$ is
 the usual mass parameter for the asymptotically free coupling defined
 at scale $\mu$:
\be\label{eq20}
\Lambda \equiv \mu e^{-{1}/{2\beta_0 g(\mu)^2}} \ ,\hspace{3em}
\beta_0 = \frac{3N_c}{16\pi^2} \ .
\ee
 As usual, $\beta_0$ denotes the first coefficient of the
 $\beta$-function and we consider here the gauge group $SU(N_c)$.

 The effective action in (\ref{eq19}) incorporates the breaking of the
 discrete $Z_{2N_c}$ chiral symmetry by the {\em gluino condensate}
\be\label{eq21}
\langle \lambda\lambda \rangle =  C\Lambda^3 e^{2\pi ik/N_c} \ .
\ee
 The phase factor depending on the integer $k$ refers to the different
 ground states defined in (\ref{eq11}).
 The proportionality factor $C$ depends, of course, on the
 renormalization scheme belonging to $\Lambda$.
 Instanton calculations and other reasonings imply that we have
 $C=32\pi^2$ in the dimensional reduction scheme
 $\Lambda = \Lambda_{DR}$ \cite{FINPOU}.

 The main assumption needed to derive the VY effective action
 (\ref{eq20}) is the choice of the chiral superfield $S$ as the
 dominant degree of freedom of low energy dynamics.
 Making a more general ansatz also containing gluon-gluon composites
 leads to a generalization and to two mixed supermultiplets in the
 low energy spectrum \cite{FAGASCH}.
 Even if $S$ is accepted as the dominant variable, one can argue about
 the existence of a chirally symmetric ground state, in addition to the
 $N_c$ ground states with broken chiral symmetry given by the integer
 $k < N_c$ in (\ref{eq21}) \cite{KOVSHIF}.

 An interesting question is how the spectrum of glueballs, gluinoballs
 and gluino-glueballs is influenced by the soft supersymmetry breaking
 due to a non-zero gluino mass $m_{\tilde{g}} \ne 0$.
 For small $m_{\tilde{g}}$ it is possible to derive the coefficients
 of the terms linear in $m_{\tilde{g}}$ in the mass formulas
 \cite{GLBMASS}.
 (Note, however, that the two papers in this reference arrive to
 different results.)

 A general consequence of the chiral symmetry breaking is the existence
 of a first order phase transition at $m_{\tilde{g}}=0$.
 At this point the different ground states in (\ref{eq21}) are
 degenerate and a coexistence of the corresponding phases is possible.
 In a mixed phase situation, as usual at first order phase transitions,
 the different phases are separated by ``bubble wall'' interfaces.
 The {\em interface tension} of the walls can be exactly derived from
 the central extension of the $N=1$ SUSY algebra \cite{KOSHSM}.
 The result is that the energy density of the interface wall is related
 to the jump of the gluino condensate by
\be\label{eq22}
\epsilon = \frac{N_c}{8\pi^2} \left| \langle \lambda\lambda \rangle_1 
- \langle \lambda\lambda \rangle_2 \right| \ .
\ee
 Combining this with eq.~(\ref{eq21}) implies that the dimensionless
 ratio $\epsilon/|\langle \lambda\lambda \rangle|$ is predicted
 independently of the renormalization scheme.

 In order to compare the predictions (\ref{eq21}) and (\ref{eq22}) to
 the results of lattice Monte Carlo simulations it is convenient to
 switch to the lattice $\Lambda$-parameter $\Lambda_{LAT}$.
 First one can use \cite{FINPOU}
\be\label{eq23}
\Lambda_{DR}/\Lambda_{\overline{MS}} = \exp\{-1/18\}
\ee
 and then for the Curci-Veneziano lattice action \cite{LAMBDALAT}
$$
\Lambda_{\overline{MS}}/\Lambda_{LAT} = \exp\left\{
-\frac{1}{\beta_0} \left[ \frac{1}{16N_c} - N_c P
+ \frac{N_c n_a}{2} P_3 \right]\right\} \ ,
$$
\be\label{eq24}
\beta_0 = \frac{N_c}{48\pi^2} (11-2n_a) \ ,\hspace{2em}
P = 0.0849780(1) \ ,\hspace{2em} P_3 = 0.0066960(1) \ .
\ee
 Here $n_a$ is the number of Majorana fermions in the adjoint
 representation, that is for SYM we set $n_a=1$.

 For transforming (\ref{eq21}) and (\ref{eq22}) to lattice units we
 need, in fact, the value of $a\Lambda_{LAT}$ at the particular values
 of interest of the lattice bare parameters $\beta,K$.
 Before performing the lattice simulations this is, of course, not
 known.
 An order of magnitude estimate can be obtained from pure gauge theory
 (at $K=0$) by noting that both for $N_c=2$ and $N_c=3$ we have for
 the lowest gluball mass $M$ \cite{GLUEBALL}
\be\label{eq25}
a\Lambda_{LAT} \simeq \frac{aM}{200} \ .
\ee
 Assuming this approximate relation also at the critical line for
 zero gluino mass, we can use eqs.~(\ref{eq21})-(\ref{eq25}) for
 estimating orders of magnitudes. 
 In the region where $aM={\cal O}(1)$ we obtain
\be\label{eq26}
a^3 |\langle \lambda\lambda \rangle_1 -
     \langle \lambda\lambda \rangle_2| = {\cal O}(1) \ ,
\hspace{3em}
a^3\epsilon = {\cal O}(10^{-1}) \ .
\ee
 As these numbers show, the predicted first order phase transition is,
 in fact, strong enough for a relatively easy observation in lattice
 simulations.

\section{Numerical Monte Carlo simulations}                 \label{sec3}

 The lattice Monte Carlo simulations of quantum field theories are
 performed in Euclidean space-time.
 For SYM, and more generally for a Yang-Mills theory of Majorana
 fermions in the adjoint representation (``gaugino'' or in the context
 of strong interactions ``gluino'') with arbitrary mass we need first of
 all the definition of Majorana fermions in Euclidean space-time.

 In the literature one may sometimes find the statement that there are
 no Euclidean Majorana spinors (see, for instance, \cite{LEUSMI}).
 This is only true as long as one is concentrating on the hermiticity
 properties of fields, as in Minkowski space.
 The definition required for an Euclidean path integral can be based on
 the appropriate analytic continuation of expectation values
 \cite{MAJORANA}.
 The essential point is that for Majorana fermions the Grassmann
 variables $\Psi$ and $\overline{\Psi}$ are not independent, as is the
 case for Dirac fermions, but are related by
\be\label{eq27}
\overline{\Psi} = \Psi^T C \ ,
\ee
 with $C$ the charge conjugation Dirac matrix.
 In fact, starting from an Euclidean Dirac fermion field represented by
 the pair $\psi,\overline{\psi}$ one can define two Majorana fermion
 fields satisfying (\ref{eq27}) by
\be\label{eq28}
\Psi^{(1)} \equiv \frac{1}{\sqrt{2}} ( \psi + C\overline{\psi}^T)
\ , \hspace{2em}
\Psi^{(2)} \equiv \frac{i}{\sqrt{2}} (-\psi + C\overline{\psi}^T) \ .
\ee
 Using also the inverse relations
\be\label{eq29}
\psi = \frac{1}{\sqrt{2}} (\Psi^{(1)} + i\Psi^{(2)})
\ , \hspace{2em}
\psi_c \equiv C\overline{\psi}^T =
\frac{1}{\sqrt{2}} (\Psi^{(1)} - i\Psi^{(2)}) \ ,
\ee
 one can easily relate expectation values of Majorana and Dirac fermion
 fields \cite{GLUINO}.

\subsection{Lattice actions and algorithms}               \label{sec3.1}
 Following Curci and Veneziano \cite{CURVEN}, we can take for the
 fermionic part of the SYM action the well known Wilson formulation.
 If the Grassmanian fermion fields in the adjoint representation are
 denoted by $\psi^r_x$ and $\overline{\psi}^r_x$, with $r$ being the
 adjoint representation index ($r=1,..,N_c^2-1$ for SU($N_c$) ), then
 the fermionic part of the lattice action is: 
\be  \label{eq30}
S_f = \sum_x \LCB \overline{\psi}_x^r\psi_x^r
-K \sum_{\mu=1}^4 \left[
\overline{\psi}_{x+\hat{\mu}}^r V_{rs,x\mu}(1+\gamma_\mu)\psi_x^s
+\overline{\psi}_x^r V_{rs,x\mu}^T (1-\gamma_\mu)
\psi_{x+\hat{\mu}}^s \right] \RCB \ .
\ee
 Here $K$ is the hopping parameter, the irrelevant Wilson parameter
 removing the fermion doublers in the continuum limit is fixed to
 $r=1$, and the matrix for the gauge-field link in the adjoint
 representation is defined as
\be  \label{eq31}
V_{rs,x\mu} \equiv V_{rs,x\mu}[U] \equiv
2 {\rm Tr}(U_{x\mu}^\dagger T_r U_{x\mu} T_s)
= V_{rs,x\mu}^* =V_{rs,x\mu}^{-1T} \ .
\ee
 The generators $T_r \equiv \half \lambda_r$ satisfy the usual
 normalization ${\rm Tr\,}(\lambda_r\lambda_s)=\half$.
 In case of SU(2) ($N_c=2$) we have $T_r \equiv \half \tau_r$ with the
 isospin Pauli-matrices $\tau_r$.
 The normalization of the fermion fields in (\ref{eq30}) is the
 usual one for numerical simulations.
 The full lattice action is the sum of the pure gauge part and
 fermionic part: 
\be  \label{eq32}
S = S_g + S_f \ .
\ee
 Here the standard Wilson action for the SU($N_c$) gauge field $S_g$
 is a sum over the plaquettes
\be  \label{eq33}
S_g  =   \beta \sum_{pl}                                                  
\left( 1 - \frac{1}{N_c} {\rm Re\,Tr\,} U_{pl} \right) \ ,   
\ee
 with the bare gauge coupling given by $\beta \equiv 2N_c/g^2$.

 Using the relations in (\ref{eq29}) one can decompose $S_f$ as a sum
 over the two Majorana components:
\be  \label{eq34}
S_f = \sum_{xu,yv} \overline{\psi}^v_y Q_{yv,xu} \psi^u_x
= \half\sum_{j=1}^2
\sum_{xu,yv} \overline{\Psi}^{(j)v}_y Q_{yv,xu} \Psi^{(j)u}_x \ ,
\ee
 where the {\em fermion matrix} $Q$ is defined in (\ref{eq30}).
 Using this, the fermionic path integral for Dirac fermions can be
 written as
\be  \label{eq35}
\int [d\overline{\psi} d\psi] e^{-S_f} = 
\int [d\overline{\psi} d\psi] e^{-\overline{\psi} Q \psi} = \det Q
= \prod_{j=1}^2 \int [d\Psi^{(j)}] 
e^{ -\half\overline{\Psi}^{(j)}Q\Psi^{(j)} } \ .
\ee
 For Majorana fields the path integral involves only $[d\Psi^{(j)}]$,
 either with $j=1$ or $j=2$.
 For $\Psi \equiv \Psi^{(1)}$ or $\Psi \equiv \Psi^{(2)}$ we have
\be  \label{eq36}
\int [d\Psi] e^{ -\half\overline{\Psi} Q \Psi }
= \pm \sqrt{\det Q} \ .
\ee

 In order to define the sign in (\ref{eq36}) one has to consider the
 {\em Pfaffian} of the antisymmetric matrix
\be  \label{eq37}
M \equiv CQ \ .
\ee
 This can be defined for a general complex antisymmetric matrix
 $M_{\alpha\beta}=-M_{\beta\alpha}$ with an even number of dimensions
 ($1 \leq \alpha,\beta \leq 2N$) by a Grassmann integral as
\be  \label{eq38}
{\rm Pf}(M) \equiv
\int [d\phi] e^{-\half\phi_\alpha M_{\alpha\beta} \phi_\beta}
= \frac{1}{N!2^N} \epsilon_{\alpha_1\beta_1 \ldots \alpha_N\beta_N}
M_{\alpha_1\beta_1} \ldots M_{\alpha_N\beta_N} \ .
\ee
 Here, of course, $[d\phi] \equiv d\phi_{2N} \ldots d\phi_1$, and 
 $\epsilon$ is the totally antisymmetric unit tensor.
 One can easily show that
\be  \label{eq39}
\left[{\rm Pf}(M)\right]^2 = \det M \ .
\ee
 If $M$ is taken from (\ref{eq37}) one also has $\det M = \det Q$.

 The relations in (\ref{eq36}) or (\ref{eq39}) show that, in order to
 represent a Majorana fermion, in the path integral over the gauge field
 the square root of the fermion determinant (or the Pfaffian of the
 matrix in (\ref{eq37})) has to be taken.
 In this sense a Majorana fermion corresponds to a flavour number
 $\half$.
 Concerning the sign of the square root, in numerical simulations it is
 easier to take always the absolute value.
 This presumably does not have an influence in the continuum limit
 because in the continuum the (real) eigenvalues of the Dirac matrix
 come in pairs and the square root is always positive (see, for
 instance, \cite{HSU}).

 In general, the lattice action describing a given ``target'' continuum
 quantum field theory is not unique.
 Besides the Curci-Veneziano action discussed up to now, another
 possibility is based on five-dimensional domain walls
 \cite{OVERLAP,DOMAIN,AONAZE}.
 In this approach one knows the value of the bare fermion mass where
 the supersymmetric continuum limit is best approached and one has
 advantages from the point of view of the speed of symmetry restoration.
 The price one has to pay is the proliferation of (auxiliary) fermion
 flavours.
 Another proposal for reaching supersymmetric quantum field theories
 is to try direct dimensional reduction on the lattice \cite{REDUCE}.

 In order to perform Monte Carlo simulations with effective flavour
 number $\half$ corresponding to Majorana fermions in the lattice
 formulation of Curci and Veneziano, one can either use the
 {\em multi-bosonic technique} \cite{LUSCHER,GLUINO} or apply the
 {\em hybrid classical dynamics algorithm} \cite{DONGUA}.
 Exploratory studies have been started recently.
 (For a recent review and status report see \cite{EDINBURGH}.)

 The first step in numerical simulations is to consider the
 {\em quenched approximation}, which neglects the dynamical effects of
 gluinos \cite{KOUMON,DOGUHEVL}.
 Since quenching breaks supersymmetry explicitly, this mainly serves
 as a testing ground for mass measurements and helps to localize the
 physically interesting bare parameter range.

 A first large scale numerical simulation of SU(2) SYM with dynamical
 gluinos has been started recently by the DESY-M\"unster collaboration
 \cite{DESYMUNSTER} using the supercomputers at HLRZ, J\"ulich and
 DESY, Zeuthen.
 The main goals of this collaboration are: to find the first order phase
 transition at zero gluino mass and to determine the masses of the
 lowest bound states formed out of gluons and gluinos in the interesting
 range of the gluino mass.

\vspace{1.5em}\noindent
{\large\bf Acknowledgements} 

\vspace{1em}\noindent
 I thank Peter Weisz for correspondence and for communicating his result
 on the ratio of $\Lambda$-parameters in (\ref{eq24}).


\end{document}